\documentclass[11pt,a4paper]{article}
\usepackage{jheppub_kim}

\usepackage{pdflscape}
\usepackage{amsmath}
\usepackage{amssymb}
\usepackage{dcolumn}
\usepackage{bm}
\usepackage{color}
\usepackage{epsfig}
\usepackage{amsfonts}
\usepackage{graphicx}
\usepackage{subfigure}
\usepackage{dcolumn}

\newcommand{\be}{\begin{equation}}
\newcommand{\ee}{\end{equation}}
\newcommand{\bea}{\begin{eqnarray}}
\newcommand{\eea}{\end{eqnarray}}

\def\be{\begin{equation}}
\def\ee{\end{equation}}
\def\bea{\begin{eqnarray}}
\def\eea{\end{eqnarray}}

\begin{document}

\title{A cosmological model of the early universe based on ECG with variable $\Lambda$-term in Lyra geometry}

\author[a]{H. Saadat,}

\affiliation[a]{Department of Physics, Shiraz Branch, Islamic Azad University, P. O. Box 71555-477, Shiraz, Iran}

\emailAdd{hsaadat2002@yahoo.com}

\abstract{In this paper, we study interacting extended Chaplygin gas as dark matter and quintessence scalar field as dark energy with an effective $\Lambda$-term in Lyra manifold. As we know Chaplygin gas behaves as dark matter at the early universe while cosmological constant at the late time. Modified field equations are given and motivation of the phenomenological models discussed in details. Four different models based on the interaction term are investigated in this work. Then, we consider other models where Extended Chaplygin gas and quintessence field play role of dark matter and dark energy respectively with two different forms of interaction between the extended Chaplygin gas and quintessence scalar field for both constant and varying $\Lambda$. Concerning to the mathematical hardness of the problems we discuss results numerically and graphically. Obtained results give us hope that proposed models can work as good models for the early universe with later stage of evolution containing accelerated expansion.}


\maketitle

\section{Introduction}
A huge effort of fundamental physics to explain the dynamics of the universe opens big window for different speculations. However, the most intriguing effect in the recent universe can not be explained properly and gives different philosophical fight between different minds. If we accept that Einstein was correct with his general relativity theory to explain accelerated expansion of the universe could be explained by negative pressure working against gravity. The belief of Einstein to the static universe made him to think about negative pressure which will stop the attraction of the gravity. However, we know that we have non static universe, moreover we know that we have accelerated expansion. According to the last observational data analysis we estimated the amount of the negative pressure in our universe, which we call it as dark energy. The simple question about the nature of the dark energy still one of the intriguing questions and left free space for new speculations. A big class of the dark energy models are introduced in life by hand and used with the field equations of general relativity. This is working but is not a satisfactory approach, therefore modification of the field equations on the Lagrangian level could be considered more fundamental and satisfactory. This approach opened a wide range of the modified general relativity, but still with a big uncertainty, because the way and the form of the modifications plays a crucial role in our understanding and depends on the form of the modification of the same class of the modified general relativity gives different forms of the dark energy. This is model dependent approach and still left a big room for speculations, of course some of the approaches could be combined with particle physics and can give a hope that this link will give us possibility to illuminate phenomenology and have a fundamental theory. Consideration of the modified general relativity promises new insight into our understanding of the universe and one of the hot and modern subject for the study. In this work we will consider one of the modifications and would like to see the behavior of the universe. Our analysis is based on exotic type of fluid which is called Chaplygin gas (CG) and sued to describe accelerating expansion of universe \cite{kamenshchik,Bento:2002ps}. This is indeed based on CG equation of state and developed to the generalized Chaplygin gas (GCG) \cite{P83}.
It is also possible to study viscosity in GCG \cite{P86,P87,P88,P90,P91,P89}.
Then, GCG was extended to the modified Chaplygin gas (MCG) \cite{P92}. Recently, viscous MCG is also suggested and studied \cite{P93,P94}. A further extension of CG model is called modified cosmic Chaplygin gas (MCCG) which was proposed recently \cite{P95,P96,P97}. The MCG equation of state (EoS) has two parts, the first term gives an ordinary fluid obeying a linear barotropic EoS, and the second term relates pressure to some power of the the inverse of energy density. However, it is possible to consider barotropic fluid with quadratic EoS or even with higher orders EoS \cite{P100-1,P100-2,P100-3}. Therefore, it is interesting to extend MCG EoS which recovers at least barotropic fluid with quadratic EoS. This is called extended Chaplygin gas (ECG) \cite{P102,P103,P104,P105,P106,P107}.\\
Another interesting models to describe dark energy are quintessence scalar field
\cite{Ratra:1987rm,Wetterich:1987fm,Khurshudyan:2014a,Liddle:1998xm,Guo:2006ab,Dutta:2009yb}, a phantom field
\cite{Caldwell:1999ew,Caldwell:2003vq,Nojiri:2003vn,Onemli:2004mb,
Saridakis:2008fy,Saridakis:2009pj,Gupta:2009kk}, the combination of both fields in a unified scenario which is called quintom \cite{Guo:2004fq,Zhao:2006mp,Cai:2009zp}, or more complicated
models such as in K-essence \cite{ArmendarizPicon:2000ah}.\\
Here, we would to combine ECG and quintessence field as the following scenario. We assume ECG as dark matter which is reasonable assumption at least in the early universe, however with possibility to extension in present epoch. The other component, dark energy, described by quintessence field. An important point is possibility of interaction between dark matter and dark energy. Cosmological models including Chaplygin gas usually used for unification of dark matter and dark energy, therefore we have a section of single ECG which can play role of both dark matter and dark energy.\\
On the other hand a unified treatment of cosmological models is given within the context of Lyra manifold \cite{Lyra01}. In that case there are some recent works of cosmology including Lyra manifold \cite{Lyra02,Lyra03} where $\Lambda$ considered as a variable. So, we construct a new cosmological model and confirm it using the recent observational data. Some advantages of this work are as follows.\\
1- First of all we use recently proposed type of Chaplygin gas.\\
2- Reducing $n+2$ free parameters of the ECG model to 3.\\
3- Interacting quintessence and ECG is a new model.\\
4- Comparison with the recent BICEP2 data.\\
5- Several types of interaction term.\\
6- Presence of Lyra manifold.\\
7- Varying $\Lambda$.\\
The paper is organized as follows. In next section we write down field equations which govern our model. Then, in section 3 we introduce the ECG and represent some cosmological behavior of the single ECG model. In section 4 we introduce our main models based on different forms of interaction term between quintessence field and ECG. In section 5 we have numerical analysis of the case of constant $\Lambda$ which extend to the case of varying $\Lambda$ in section 6. Finally in section 7 we summarize our results and give conclusion.

\section{Field equations}
Field equations that govern our model are given by \cite{Lyra02},
\begin{equation}\label{eq:Einstein eq}
R_{\mu\nu}-\frac{1}{2}g_{\mu\nu}R-\Lambda g_{\mu \nu}+\frac{3}{2}\varphi_{\mu}\varphi_{\nu}-\frac{3}{4}g_{\mu \nu}\varphi^{\alpha}\varphi_{\alpha}=T_{\mu\nu},
\end{equation}
where $\varphi_{\mu}$ is a characteristic of Lyra geometry.
Considering the content of the universe to be a perfect fluid, we have,
\begin{equation}\label{eq:T}
T_{\mu\nu}=(\rho+P)u_{\mu}u_{\nu}-Pg_{\mu \nu},
\end{equation}
where $u_{\mu}=(1,0,0,0)$ is a 4-velocity of the co-moving
observer, satisfying $u_{\mu}u^{\mu}=1$. Let $\varphi_{\mu}$ be a time-like
vector field of displacement,
\begin{equation}
\varphi_{\mu}=\left ( \frac{2}{\sqrt{3}}\beta,0,0,0 \right ),
\end{equation}
where $\beta=\beta(t)$ is a function of time alone, and the factor $\frac{2}{\sqrt{3}}$ is substituted in order to simplify the writing of all the following equations.
By using FRW metric for a flat universe,
\begin{equation}\label{s2}
ds^2=-dt^2+a(t)^2\left(dr^{2}+r^{2}(d\theta^{2}+\sin^{2}\theta d\phi^{2})\right),
\end{equation}
field equations can be reduced to the following Friedmann equations,
\begin{equation}\label{eq:f1}
3H^{2}-\beta^{2}=\rho+\Lambda,
\end{equation}
and,
\begin{equation}\label{eq:Freidmann2}
2\dot{H}+3H^{2}+\beta^{2}=-P+\Lambda,
\end{equation}
where $H=\dot{a}/{a}$ is the Hubble parameter, and an over dot
stands for differentiation with respect to cosmic
time $t$, and $a(t)$ represents the scale factor.\\
The continuity equation reads as,
\begin{equation}\label{eq:coneq}
\dot{\rho}+\dot{\Lambda}+2\beta\dot{\beta}+3H(\rho+P+2\beta^{2})=0.
\end{equation}
With an assumption that,
\begin{equation}\label{eq:DEDM}
\dot{\rho}+3H(\rho+P)=0.
\end{equation}
The equation (\ref{eq:coneq}) will give a link between $\Lambda$ and $\beta$ of the following form,
\begin{equation}\label{eq:lbeta}
\dot{\Lambda}+2\beta\dot{\beta}+6H\beta^{2}=0.
\end{equation}
In order to introduce an interaction between the dark energy and dark matter, we should mathematically split the equation (\ref{eq:DEDM}) into the following equations,
\begin{equation}\label{eq:inteqm}
\dot{\rho}_{DM}+3H(\rho_{DM}+P_{DM})=Q,
\end{equation}
and,
\begin{equation}\label{eq:inteqG}
\dot{\rho}_{DE}+3H(\rho_{DE}+P_{DE})=-Q,
\end{equation}
where $Q$ is interaction term and expressed explicitly in the section 4. For the Dark energy sector we assume quintessence field which also introduce in the section 4. Before that we introduce the extended Chaplygin gas in the next section.
\section{Extended Chaplygin gas}
In this section we review evolution of CG EoS which is given by the following primary relation,
\begin{equation}\label{I2}
P_{CG}=-\frac{B}{\rho_{CG}},
\end{equation}
where $B$ is an arbitrary parameter which usually considered as a constant.\\
In order for the universe passes from a dust dominated epoch to a de
Sitter phase through an intermediate phase described as mixture of cosmological
constant and radiation the Chaplygin gas equation of state generalized to \cite{EPJC73(2013)2295},
\begin{equation}\label{I3}
P_{GCG}=-\frac{B}{\rho_{GCG}^{\alpha}},
\end{equation}
where $\alpha$ and $B$ are free parameters. It is clear that $\alpha=1$ reproduces the pure Chaplygin gas model given by (\ref{I2}). This model is called the generalized Chaplygin gas model. There is also a class of equations of state that interpolate between
standard fluids at high energy densities and generalized Chaplygin gas fluids at low energy densities which is called the modified Chaplygin
gas with the following equation of state \cite{P83 1211.3518},
\begin{equation}\label{I5}
P_{MCG}=A\rho_{MCG}-\frac{B}{\rho_{MCG}^{\alpha}},
\end{equation}
where $A$, $\alpha$, and $B$ are free parameters of the model. The case of $A = 0$ recovers generalized Chaplygin gas equation of state, and $A = 0$ together $\alpha = 1$ recovers the original Chaplygin gas equation of state.
The MCG equation of state has two parts, the first term gives an ordinary fluid obeying a linear barotropic equation of state, while there are some models with the quadratic equation of state \cite{P100-1,P100-2,P100-3},
\begin{equation}\label{I1}
P = P_{0}+\omega_{1}\rho+\omega_{2}\rho^{2},
\end{equation}
where $P_{0}$, $\omega_{1}$ and $\omega_{2}$ are constants. Easily we can set $P_{0}=\omega_{2}=0$  to recover linear barotorpic equation of state.\\
Modified Chaplygin gas include only barotropic fluid with linear equation of state, while it is possible to extend them to including quadratic barotropic equation of state given by (\ref{I1}). It yields to introducing extended Chaplygin gas with the following equation of state,
\begin{equation}
P_{ECG}=\sum_{i=1}^{n}{A_{i}\rho_{ECG}^{i}}-\frac{B}{\rho_{ECG}^{\alpha}},
\end{equation}
where $\alpha$ and $B$ are constants. It is obvious that the $n=1$ reduced to the modified Chaplygin gas with $A=1/2$. Moreover,
\begin{equation}\label{eq:EoSCh}
\omega_{ECG}=\frac{P_{ECG}}{\rho_{ECG}}=\sum_{i=1}^{n}{A_{i}\rho_{ECG}^{i-1}}-\frac{B}{\rho_{ECG}^{1+\alpha}},
\end{equation}
is EoS of ECG, and,
\begin{equation}\label{s35}
C_{s}^{2}=\frac{dP_{ECG}}{d\rho_{ECG}},
\end{equation}
is sound speed. It is one of the important point of this paper which relate the coefficient of $\rho_{ECG}^{n}$ to $n$, so we reduced $n$ free parameters of the model to one.\\
In the next section we consider ECG as dark matter and assume quintessence field as dark energy to construct an interacting model.
\section{The models}
We assume quintessence scalar field as dark energy described by a scalar field $\phi$ and self interacting potential $V(\phi)$. Energy density and pressure given by,
\begin{equation}\label{eq:rhoQ}
\rho_{\phi}=\frac{1}{2}\dot{\phi}^{2}+V(\phi),
\end{equation}
and,
\begin{equation}\label{eq:rhoP}
P_{\phi}=\frac{1}{2}\dot{\phi}^{2}-V(\phi).
\end{equation}
We consider models for the universe where an effective energy density and pressure given by,
\begin{equation}\label{eq:rhoeff}
\rho=\rho_{\phi}+\rho_{ECG},
\end{equation}
and,
\begin{equation}\label{eq:Peff}
P=P_{\phi}+P_{ECG},
\end{equation}
where $\rho_{\phi}$ and $P_{\phi}$ are energy density and pressure of the quintessence dark energy in the universe, while $\rho_{ECG}$ and $P_{ECG}$ represent energy density and pressure of the ECG. An important point is possibility of interaction between components.\\
The solving strategy and structure of the problem allow us to assume that the form of the potential $V(\phi)$ is given as follow,
\begin{equation}\label{eq:potV}
V(\phi)=V_{0}exp{\left [-\frac{A}{2}\phi^{\gamma} \right ]},
\end{equation}
where $A$ and $\gamma$ are arbitrary constants. Moreover, to make the system of the differential equation closed we assume two different situations, namely once we consider $\Lambda$ to be a constant, while for the second case $\Lambda$ is a dynamical quantity and we consider one of the classical form,
\begin{equation}\label{Lambda}
\Lambda=\rho=\rho_{\phi}+\rho_{ECG}.
\end{equation}
In this work we will consider three different types of interaction term \cite{Interaction_Saridakis} which gives us six different models depend on varying or constant $\Lambda$. At the first step we will perform numerical analysis of the important cosmological parameters in the case of constant $\Lambda$ then consider varying $\Lambda$. Different forms of interesting interaction terms between quintessence and ECG provide us a huge number of the models. We start with the interaction term $Q$ given by,
\begin{equation}\label{eq:Q1}
Q=3Hb\rho_{\phi},
\end{equation}
then we consider the second model where $Q$ reads as \cite{1410.5858},
\begin{equation}\label{eq:Q3}
Q=H^{\lambda}b\rho_{ECG}^{m},
\end{equation}
where $b$, $\lambda$ and $m$ are arbitrary constants. So, the parameter $b$ can be interpreted as strength of interaction. The case of $b=0$ gives non-interacting model.\\
We start analysis of the models from the next section and start with the case where $\Lambda$ is a constant
\section{Cases of constant $\Lambda$}
Consideration of the constant $\Lambda$ can simplify field equations, and before graphical analysis we would like to provide mathematical forms of the simplified field equations. According to this assumption, the equation given by (\ref{eq:coneq}) will be modified as,
\begin{equation}
\dot{\rho}+2\beta\dot{\beta}+3H(\rho+P+2\beta^{2})=0.
\end{equation}
So, using the equation (\ref{eq:DEDM}) we have,
\begin{equation}
\dot{\beta}+3H\beta=0.
\end{equation}
An integration of the last equation reveals dependence between $\beta$ and the scale factor $a$ as follow,
\begin{equation}
\beta=\frac{a_{0}}{a^{3}},
\end{equation}
where $a_{0}$ is an integration constant.
\subsection{Model 1}
\begin{figure}[h!]
 \begin{center}$
 \begin{array}{cccc}
\includegraphics[width=60 mm]{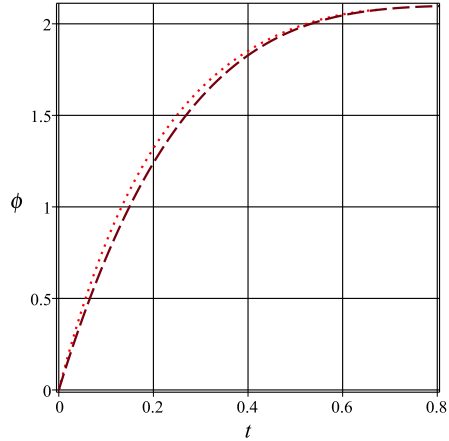}
 \end{array}$
 \end{center}
\caption{Typical behavior of the scalar field against $t$ ($10^{17}\hspace{1mm}s$) for the model 1. Dashed line represent behavior of the function given by the equation (\ref{M13}), while dotted line represent numerical solution of the equation (\ref{M11}).}
 \label{1}
\end{figure}

\begin{figure}[h!]
 \begin{center}$
 \begin{array}{cccc}
\includegraphics[width=60 mm]{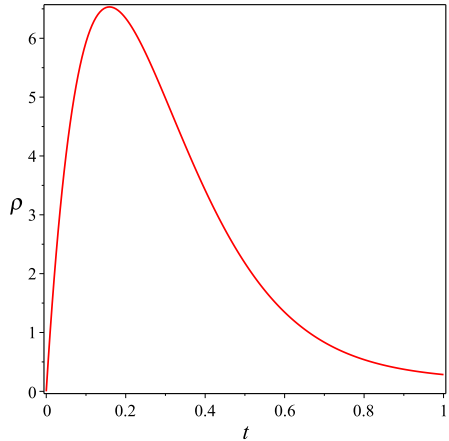}
 \end{array}$
 \end{center}
\caption{Typical behavior of the $\rho_{ECG}$ against $t$ ($10^{17}\hspace{1mm}s$) for the model 1.}
 \label{2}
\end{figure}

The first model has the interaction term given by the equation (\ref{eq:Q1}) with the constant $\Lambda$ which yields to the following equations,
\begin{equation}\label{M11}
\ddot{\phi}+3H(1+\frac{b}{2})\dot{\phi}+\frac{3HbV(\phi)}{\dot{\phi}}+\frac{dV(\phi)}{d\phi}=0,
\end{equation}
and,
\begin{equation}\label{M12}
\dot{\rho}_{ECG}+3H(1+\omega_{ECG})\rho_{ECG}-3Hb(\frac{1}{2}\dot{\phi}^{2}+V(\phi))=0,
\end{equation}
where $\omega_{ECG}$ is the EoS parameter of the ECG given by (\ref{eq:EoSCh}).\\
We solve the equation (\ref{M11}) numerically and fit the following function of the scalar field,
\begin{equation}\label{M13}
\phi=2.1(1-(x-1)^4),
\end{equation}
Numerical result and best fitted value drawn in the Fig. \ref{1}. We can see that the scalar field is increasing function of time and yields to a constant value at the late time.\\ Then, we solve the equation (\ref{M12}) numerically and find typical behavior of the extended Chaplygin gas density which illustrated in the Fig. \ref{2}. We can see interesting behavior, the energy density increases first to reach maximum value and then decrease to the infinitesimal value corresponding to the current value.

\begin{figure}[h!]
 \begin{center}$
 \begin{array}{cccc}
\includegraphics[width=60 mm]{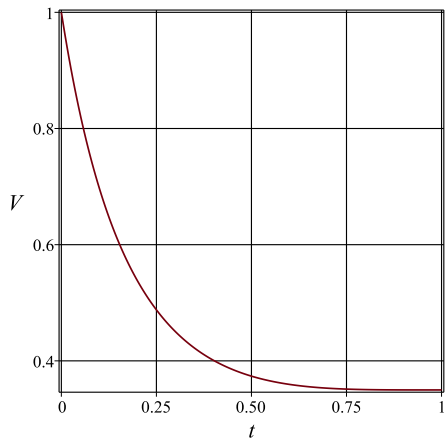}
 \end{array}$
 \end{center}
\caption{Typical behavior of the scalar potential against $t$ ($10^{17}\hspace{1mm}s$) for the model 1.}
 \label{3}
\end{figure}

We find good result corresponding to the case of $n=3$ which is good news for our model which is a confirmation of the ECG, at least to the third order. Moreover, we can see far results away than observations for the case of $b=0$ and $n>1$, which means presence of interaction is necessary in the ECG model.\\
Consideration of interaction terms bring us to investigate two important parameters $\Omega_{\phi}=\rho_{\phi}/3H^{2}$ and $\Omega_{ECG}=\rho_{ECG}/3H^{2}$ to see solving coincident problem of the cosmological constant model. In our model, which cosmological constant is presence, we can check and find values of $\Omega_{\phi}$ and $\Omega_{ECG}$ at the same order.\\
We also able to investigate evolution of the scalar potential of the model which presented by the Fig. \ref{3}. We can see that the scalar potential is decreasing function of time and vanishes at the late time as expected.

\subsection{Model 2}
The second model obtained using the interaction term given by the equation (\ref{eq:Q3}) and gives two following differential equations,
\begin{equation}
\ddot{\phi}+3H\dot{\phi}+H^{\lambda}b\frac{\rho_{ECG}^{m}}{\dot{\phi}}+\frac{dV(\phi)}{d\phi}=0,
\end{equation}
and,
\begin{equation}
\dot{\rho}_{ECG}+3H(1+\omega_{ECG})\rho_{ECG}-H^{\lambda}b\frac{\rho_{ECG}^{m}}{\dot{\phi}}=0,
\end{equation}
which describe the dynamics of the content of the universe. $\omega_{ECG}$ is the EoS parameter for the extended Chaplygin gas. Here, we examined higher order terms of the ECG and find good behavior of the Hubble and deceleration parameters. Analysis of total EoS parameter tells that $\omega_{tot}\rightarrow-1$ at the late time for higher order terms of ECG. It means that Higher order terms are necessary.\\
Moreover, we find that values of  $\Omega_{ECG}$ and  $\Omega_{\phi}$ are of the same order. Finally we can obtain time evolution of the scalar field and scalar potential similar to the previous models.

\section{Cases of varying $\Lambda$}
Among the numerous possibilities concerning to the phenomenological form of $\Lambda$ in this section we will consider one of the classical form given by the equation (\ref{Lambda}) proposed and considered for a long time. The analysis of the first model of this section with $\Lambda$ starts by the interaction $Q=3Hb\rho_{Q}$ between two components. Mathematical formulation describing the dynamics of the energy densities of the components is the same as for the first two models, with one difference that the form of $\beta$, therefore the form of the Hubble parameter, is different. In next subsections we will provide and discuss behavior of the cosmological parameters.

\subsection{Model 3}
This model given by the equations (\ref{eq:Q1}) and (\ref{Lambda}). We find that the Hubble expansion parameter is decreasing function of time and yields to a constant at the late time which is in agreement with observations. Study of the deceleration parameter shoe acceleration to deceleration phase transition at the near past. The deceleration parameter begin with positive value (acceleration phase) and yields to minus one at the late time. Current value of this parameter $q\approx-0.8$ is available in this model.\\
Also behavior of the EoS parameter is completely expected which yields to minus one at the late time. As before, we can check and find values of $\Omega_{\phi}$ and $\Omega_{ECG}$ at the same order. Like to the previous models we find scalar potential and scalar field decreasing and increasing function of time respectively.\\
Unfortunately, stability analysis of this model based on squared sound speed show that the model is completely unstable and it bring us to consider the next model.

\subsection{Model 4}
\begin{figure}[h!]
 \begin{center}$
 \begin{array}{cccc}
\includegraphics[width=60 mm]{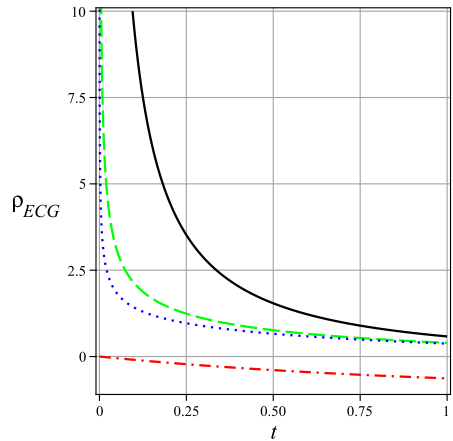}
 \end{array}$
 \end{center}
\caption{Typical behavior of the $\rho_{ECG}$ against $t$ ($10^{17}\hspace{1mm}s$) for the model 4. $m=0$ (dash dot red), $m=2$ (solid black), $m=3$ (dash green), $m=4$ (dot blue).}
 \label{4}
\end{figure}

\begin{figure}[h!]
 \begin{center}$
 \begin{array}{cccc}
\includegraphics[width=60 mm]{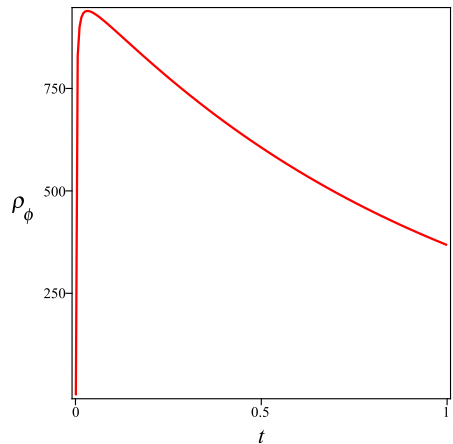}
 \end{array}$
 \end{center}
\caption{Typical behavior of the $\rho_{\phi}$ against $t$ ($10^{17}\hspace{1mm}s$) with $m=2$ for the model 4.}
 \label{5}
\end{figure}

This model given by the equations (\ref{eq:Q3}) and (\ref{Lambda}). Behavior of cosmological parameters of this model is approximately similar to the previous model but we find that this model is stable at least before the late time. So, it can describe early and current state of universe.\\
Having $P_{\phi}$ and $\rho_{\phi}$ give us scalar field and potential using the equations (\ref{eq:rhoQ}) and (\ref{eq:rhoP}). So, first we should solve the equation (\ref{eq:inteqG}) to find $\rho_{ECG}$ then use it in the equation (\ref{eq:inteqm}) to obtain scalar energy density. Therefore we can obtain evolution of $\beta$, Hubble and deceleration parameters.\\
The master equation is,
\begin{equation}
\dot{\rho}_{ECG}+3H(1+\omega_{ECG})\rho_{ECG}+H^{\lambda}b\rho_{ECG}^{m}=0.
\end{equation}
In the Fig. \ref{4} we draw typical behavior of the ECG energy density for various values of $m$. We can see that the condition $m>1$ is necessary. Also we find that evolution of $\rho_{ECG}$ for higher $m$ is faster than that with lower $m$. It is obvious that $\rho_{ECG}$ behaves like some power of exponential so it help us to find typical behavior of $\rho_{\phi}$ (see Fig. \ref{5}). We plot $\rho_{\phi}$ for $m=2$ and find that higher $m$ decreases initial values of density dramatically. However, at present we can say that $\Omega_{\phi}$ and $\Omega_{ECG}$ are at the same order.\\
Now, using the relation,
\begin{equation}
C_{s}^{2}=\frac{\dot{P}}{\dot{\rho}}\geq0,
\end{equation}
where $C_{s}$ is sound speed, it is easy to check stability of the model, we find that the model is stable at least from beginning to the late time. 

\section{Conclusion}
In this paper we proposed a new cosmological model to describe evolution of universe. Our model is based on new proposed extended Chaplygin gas. In order to have a comprehensive model we considered the case of varying $\Lambda$ in Lyra manifold. We studied also cases of constant $\Lambda$ and compare our results with the case of varying $\Lambda$ to find varying model as better case. We considered two special kind of interaction terms and concluded that the second kind yields to a good model. We found that an interacting ECG model with $Q=H^{\lambda}b\rho_{ECG}^{m}$ and $\Lambda=\rho$ is an stable model which can describe evolution of the universe.

\end{document}